# Adaptive resources allocation at the cell border using cooperative technique

Abbass Marouni, Youssef Nasser, Maryline Hélard, et Haidar El-Mokdad.
1 : Institut d'Électronique et de Télécommunications de Rennes (IETR, UMR CNRS 6164)
    INSA, 20 Avenue des Buttes de Coësmes, 35043 Rennes, France.

Contact : Marouni.Abbass@gmail.com

**Résumé**
Les techniques de communication coopérative ont connu récemment un large intérêt dans la communauté de recherche vu les grands avantages qu'elles pourraient apporter en termes de performance et de débit. Le principe de base des communications est fondé sur l'utilisation d'un relai entre la source (émetteur) et la destination (récepteur). Dans ce papier, on propose une adaptation de l'allocation des ressources dans les relais afin de profiter au maximum de ces techniques. Plus particulièrement, on est intéressé à la situation des mobiles en bordure de cellule dans un contexte multicellulaire. Dans ce contexte, les mobiles pourraient demander plusieurs types de services avec plusieurs priorités mais avec des conditions difficiles du canal de transmission. En utilisant une allocation adaptative de ressources et une optimisation intercouche, on démontre qu'on peut assurer les contraintes de priorité et de performance (application temps réel) requises par les mobiles en bordure de cellule.

**Abstract**
The technique of cooperative communication has recently gained momentum in the research community; this technique utilizes the notion of relay, as an intermediate node between the source and the destination, to enhance the overall system performance. In this paper we explored the benefits of adaptive cooperation, in which the relay adapts its relaying process in response to channel conditions and data priorities. We are particularly interested in applying this concept to the cell border situation, in which two mobile nodes acting as destinations communicate with base stations (sources) through a relay. The adaptive cooperation is proposed here since the transmission channel conditions (Packet Error Rate for example) and data priorities are not the same for both mobiles. We show that using the adaptive resource allocation technique in combination with the cross layer design techniques, we can achieve Real-Time data constraints with no additional overhead.

**Mots-clés:** Coopération, Allocation de ressource, protocole MAC, Cross layer.

## 1. Introduction

### *1.1 Cooperative Communication*

Cooperative communication technique exploits the broadcast nature of the wireless channel, in a way that a signal from a source node to a destination one can be overheard by other nodes. These nodes called *relays*, *partners* or *helpers* process the signal they overhear and then transmit the signal towards the destination [1] (FIG. 1). The processing done by the relay can range from simple forwarding to sophisticated strategies that include decoding then re-encoding, partial forwarding, and compression.

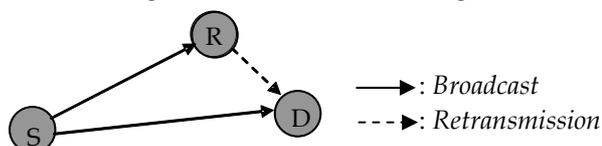

FIG. 1 - The basic idea of cooperation is to let another node retransmit an overheard packet.

Cooperative relaying can substantially decrease the error rate of a wireless transmission by using two hops link (Source - Relay, Relay - Destination) instead of a weak one hop direct link (Source – Destination). This provides higher throughput and lower delays with fewer retransmissions and adaptability to network conditions.

### *1.2 Cooperation at the physical and MAC layers*

The initial attempts for developing cooperative communications focused on the physical layer. These attempts have been interpreted in the form of new modulation and coding techniques that increased the overall performance and the throughput of the wireless system.

To realize fully cooperative networks, the notion of cooperation was extended to the MAC layer a cross layer approach was followed where information form the physical layer was used at the MAC layer that regulated the cooperation process [1].

Several publications were made on the cooperative MAC protocols and some of these protocols were implemented on real systems [2].

The innovation of cooperative communications is not confined only to the physical and MAC layers. It is available in various forms at different higher protocol layers.

### *1.3 Adaptive Resources Allocation*

The quality of wireless links suffers from time-varying channel degradations, and most of the wireless systems are limited in their ability to adapt to these channels variations because they are designed with fixed values for most system parameters. The values of these parameters are usually a compromise between the requirements for worst case channel requirements and the need for low implementation cost [3]. Adaptation can be performed at all layers of the protocol stack to accommodate the dynamics of wireless channels. In literature, we find a lot of studies on the adaptations at the MAC and physical layers since they can provide QoS guarantees made to multimedia applications [2,3]. Results have shown that adapting the MAC layer packet schedule by deferring transmission, when radio channel is in fade significantly improves link efficiency. Also the continuous adaptation of link-layer error control and packet length of the sender, in response to feedback from the radio receiver at the destination, can significantly improve the energy efficiency of data transport [3].

### *1.4 System Model*

In this paper we tackled the problem of poor wireless channels connecting mobiles (destinations), located at the cell border, to base stations (sources) by introducing a relay node between the sources and the mobiles (FIG. 2). The relay will connect sources to mobiles and will control the relaying (forwarding) of the packets in a way to adapt to channels conditions and QoS criteria imposed on data types.

Our target is to design a basic cooperative MAC protocol, that uses information from the physical layer in order to adapt its functions. It is a cross layer approach which optimises the use of the relay in poor channel conditions.

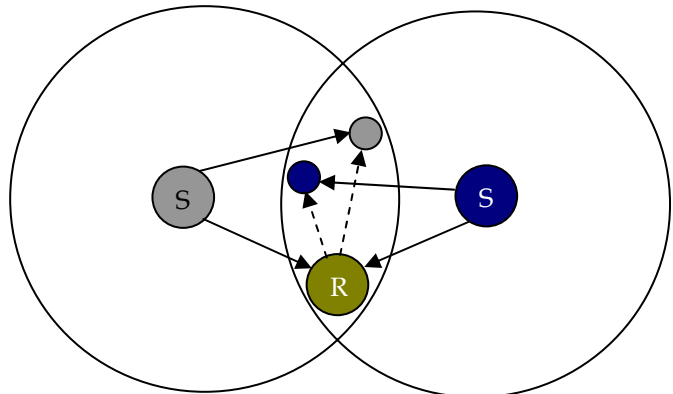

FIG. 2 – System's model with sources, destinations and the relay.

The rest of the paper is organized as follows. *Section 2* introduces the idea of cooperation at the cell border. The allocation algorithm is formulated in *Section 3* and the simulation results are given in *Section 4*. The paper's conclusion is drawn in *Section 5*.

**2. Cooperation at the cell border**

The main problem for mobiles, located at the cell border is that these mobiles receive a weak signal from the source, due to the poor coverage provided by the base station at the cell border. This weak direct link results in poor performance, longer delays and higher packet error rates. The introduction of the relay at the cell border, between two adjacent cells can improve the performance, in a way that the relay can capture the signals from the sources intended to the mobiles and adaptively retransmit the signals to the mobiles. The relay used can be a dedicated device that is installed at the cell border between two adjacent cells, or simply a mobile device that is willing to act as relay and help others.

From the cellular network model perspective, adding a relay capability to a cellular network can increase system service coverage and can also increase system capacity. Then, the system performance is enhanced with an alternative relaying capability to communicate with the base station [4]. Indeed, mobile terminals with poor signal reception can benefit from the alternative path; this could provide greater throughput or better quality of service (QoS).

Similarly, replacing a long-range high-power transmission with 2 low-power relay transmission could reduce energy consumption for mobile terminals.

In our system model, the model was presented with one single relay (FIG. 2), but the system could have more than one single relay, so we had to answer the question concerning the optimal number of relays to use (FIG. 3).

To answer this question and find the optimal number of relays between a source and a destination we setup a simulation environment using OMNET++, where we defined a source node, a destination node and a number of relay nodes between the two, then we implemented Coop-MAC protocol [2] at the source node (S). In this protocol, the source chooses the best route to send data intended to the destination from a set of routes selected by passing through relays or through the direct link.

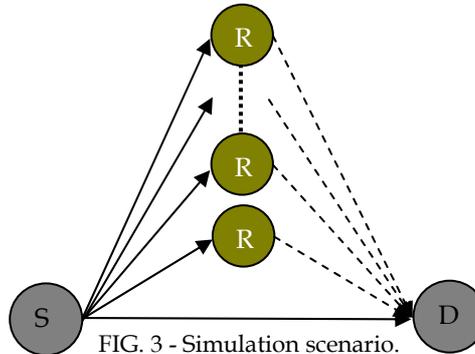

FIG. 3 - Simulation scenario.

The source in the simulation works in the same way as in CoopMAC, where the source access a special table that contains every possible route (direct or through relay) to the destination with the data rate of each route. The source computes the time needed to transmit a data packet through each route and chooses the route that gives the shortest delay [2].

The result of the simulation is given in FIG. 4, the simulation settings used are given in TAB. 1. The data rates in our model are taken from IEEE 802.11b standard [5]. The CoopMAC implemented here is a modified version of IEEE 802.11b MAC protocol [2].

| *Packet Size* | 3 KB |
|---|---|
| *S-D data rate* | *Uniformly selected between 1 and 11 mbps* |
| *S-R data rate* | *Uniformly selected between 5 and 11 mbps* |
| *R-D data rate* | *Uniformly selected between 5 and 11 mbps* |

TAB. 1 - Simulation Settings.

FIG. 4 gives the variation of the throughput in function of the number of relays. It shows that the throughput increases as the number of relays increase. This increase turns into steady convergence as the number of relays exceeds four. By adding one relay the throughput increases by 22%, from the case of direct communication without a relay, adding another relay the throughput increases by 9%, from the case of one relay, and so on. The throughput gained comes with increased overhead associated with management of the relays, and continuous update of the tables to reflect any new changes in the network.

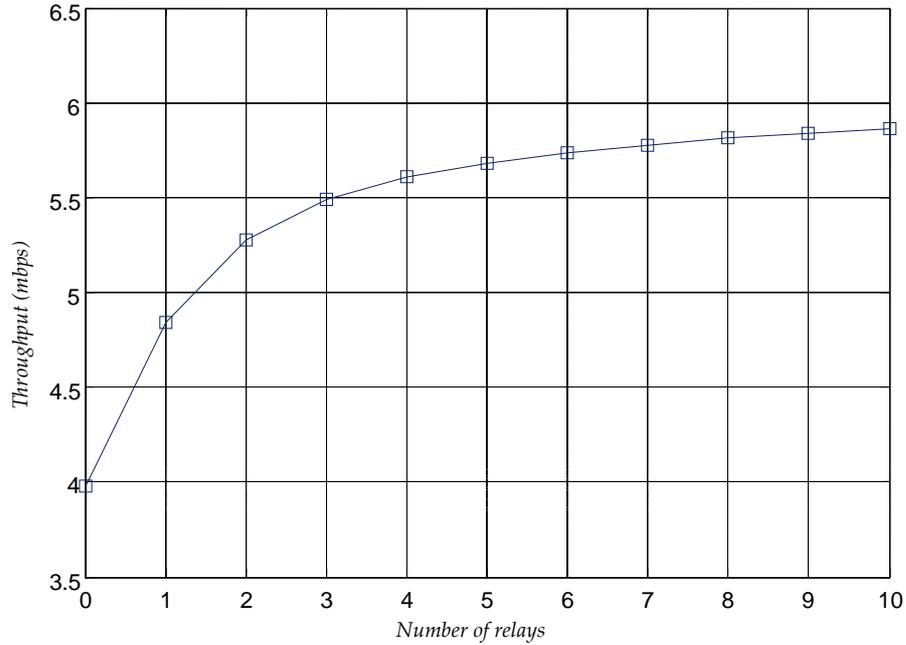

FIG. 4 - Variation of throughput with the number of relays.

The optimal number of relays to be used depends on the application, FIG. 4 shows that the throughput gained for using more than four relays is small compared to the case of less than four relays. The optimal number of relays, to be used, as deduced from FIG. 4, is four relays, but in this paper we will limit the number of relays to one relay. The results with more than one relay will be given in the final version of this paper.

With this information in hand, we can continue towards improving the performance of mobile devices, located at the border of a cell.

**3. Resource Allocation using cooperative technique**

In the previous section we showed that the use of one relay will provide us with a notable throughput gain, compared to the non-cooperative case. So one relay will be placed at the cell border (FIG. 2).

In this section, we will formulate the problem and present our solution.

*Adaptive resources allocation at the cell border using cooperative technique*

Looking back at the system model in FIG. 2, the mobile devices located at the cell border will receive a poor signal from the source directly. The source can figure out that it can reach the destination through the relay, which will provide a better signal and shorter delays. This solves the first part of the problem, and provides the source with a cooperative route to use.

The second part of the problem lies in the allocation of the relay's resources, these resources must be shared between the two mobiles, so some parameters must be used in the allocation process at the relay. In a real system scenario, we assume that the relay is allowed to receive an estimation of the channel state information or some parameter from the physical layer like the packet error rate (PER) or the signal to noise ratio (SNR). In our work we assume that the relay receives an estimation of the PER and this information will be used in combination with priorities at the MAC layer to distinguish data types, and to calculate sharing ratios. Following this cross layer approach, less parameters are involved and a rapid and better allocation is provided.

We distinguished two main data types, Real-Time (voice and video calls, video and audio streaming, etc) and non-Real-Time (data transfer, web browsing, etc), each type is associated with a target PER (TAB. 2 & FIG. 5) that is sent by the mobiles to the relay; the relay uses these target PERs in addition to an estimation of the current link's PER (FIG. 5), sent by the mobile with acknowledgment packets, to calculate the sharing ratios and allocate resources accordingly, we assume in this work that the packets sent by the sources are successfully received by the relay.

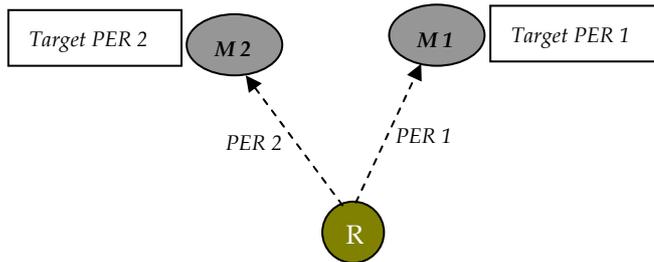

| | PER Interval = [0.0001, 0.001] |
| --- | --- |
| | Target PER 1 ∈ [0.0001, 0.001] |
| | Target PER 2 ∈ [0.0001, 0.001] |
| | Target PER (Real-Time) > Target PER (non Real-Time) |

| Application | Reliability | Delay | Bandwidth |
| --- | --- | --- | --- |
| *Real-Time* | Low | Low | High |
| *Non-Real-Time* | High | Medium | Medium |

FIG. 5 - Target and link PER.  TAB. 2 – Real-Time and non-Real-Time.

After calculating the sharing ratio of each mobile, the relay starts receiving data packets from both sources, then constructs a burst of length K that contains packets form both sources (FIG. 6). The number of packets sent by the relay for each mobile is determined from the sharing ratio that is previously calculated. The burst is then transmitted packet by packet, starting by the packets of the mobile with higher priority. The length of the burst containing data for both mobiles could be then a parameter of the simulation.

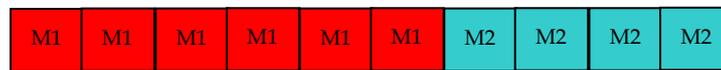

FIG. 6 - burst constructed by the relay. (Mobile 1 and 2).

### *3.1 Allocation Process*

A burst of length K is to be shared between two mobiles. The relay receives the estimations of both PERs of both mobiles and achieves the sharing process according to data priorities of each mobile and the corresponding PER targets ($PER_{T1}$, $PER_{T2}$).

**Input:** $PER_{T1}$, $PER_{T2}$, $PER_1$, $PER_2$, K *(Burst Length)*.
**Output:** $N_1$ *(Number of packets reserved for mobile 1)*, $N_2$ *(Number of packets reserved for mobile 2)*.

The target PERs, of the mobiles are changed when the mobiles require a new data type with a new Target PER. The estimations of the links PERs are received with each mobile acknowledgment after the end of each burst. Our problem is how to increase the throughput of each mobile while maintaining a target PER.

The problem could be formulated as:

$$\begin{cases} \max(N_1 + N_2) \\ N_1.PER_1 < PER_{T1} \\ N_2.PER_2 < PER_{T2} \\ N_1 + N_2 < K \end{cases} \quad \text{Constraints}$$

To solve these equations and then compute $N_1$ and $N_2$, we are going to use Lagrange optimization. The Lagrange function L is given by:

$$L = (N_1 + N_2) + \alpha(N_1.PER_1 - PER_{T_1}) + \beta(N_2.PER_2 - PER_{T_2}) + \gamma(N_1 + N_2 - K)$$

with $\alpha, \beta, \gamma$ being Lagrange multipliers

Derivatives set to zero:

$$\begin{cases} \dfrac{\delta L}{\delta N_1} = 1 + \alpha.PER_1 = 0 & (1) \\[6pt] \dfrac{\delta L}{\delta N_2} = 1 + \beta.PER_2 = 0 & (2) \\[6pt] \dfrac{\delta L}{\delta \alpha} = N_1.PER_1 - PER_{T1} = 0 & (3) \\[6pt] \dfrac{\delta L}{\delta \beta} = N_2.PER_2 - PER_{T2} = 0 & (4) \\[6pt] \dfrac{\delta L}{\delta \gamma} = N_1 + N_2 - K = 0 & (5) \end{cases}$$

We defined $a = \dfrac{PER_{T2}}{PER_{T1}}$ (6)

Using equations (3), (4), (5) and (6) we calculated N1 and N2:

$$N_1 = K \frac{PER_2}{(a.PER_1 + PER_2)} \; ; \quad N_2 = K \frac{a.PER_1}{(a.PER_1 + PER_2)}$$

## 4. Simulation Results

In this section, we present the simulation results for our proposed allocation scheme and compare it to the case of uniform non-adaptive scheme ($N_1 = N_2 = K/2$). The simulation was performed using OMNET++, the mobile nodes were implemented as generic nodes with no specified standard, while the relay node was implemented with a modified CoopMAC protocol at the MAC layer. The physical layer specifications are given in TAB. 3.

| Channel Model: Additive white Gaussian noise model |
| Coding: Convolution coding (Coding Rate 1/2) |
| Modulation Technique: 16-QAM |

TAB. 3 – Physical layer specifications

*Adaptive resources allocation at the cell border using cooperative technique*

FIG. 7 & 8, show the variations of the Throughput and PER in function of the ratio a in case of adaptive allocation and the case of uniform non-adaptive allocation. The ratio a varies between 0.25 and 4, passing by 1 which is the case when the two mobiles have the same target PER, and they are expecting the same data type. The simulations were performed with K = 10.
In FIG. 7 & 8, we changed the target PER of both mobiles and recorded the throughput and PER of each mobile in addition to the throughput and PER in the uniform case.

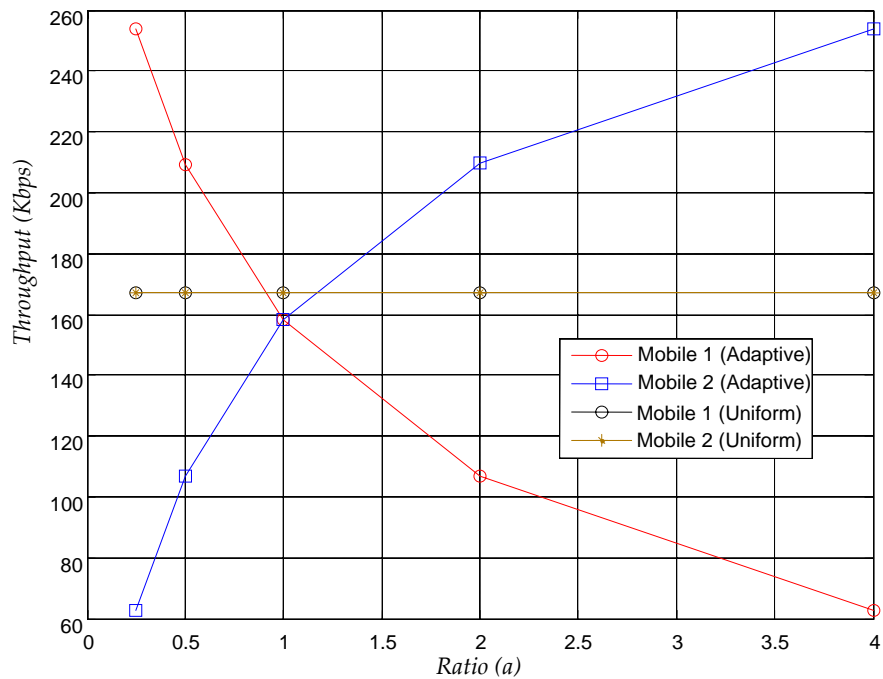
FIG. 6 – Variation of throughput in function of ratio (a).

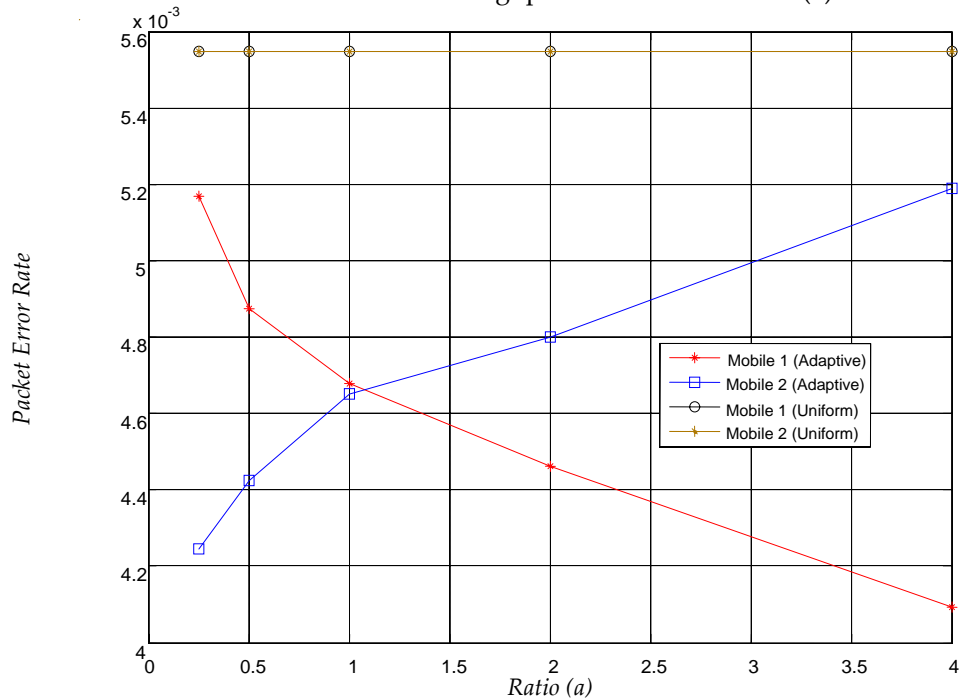
FIG. 8 - Variation of PER in function of the ratio (a).

FIG. 7 & 8 show that the Mobile with a Real-Time data (higher target PER) will get higher throughput (bandwidth) but with higher PER, the other mobile with non-Real-Time data will get lower throughput and lower PER. Of course, we can see that the PER targets are almost verified in all the cases. This compromise between throughput and PER helps achieving the criteria of TAB. 2. Moreover FIG. 8, shows that our allocation scheme outperforms the uniform non-adaptive case, by ensuring a lower PER for all values of the parameter *a*, this provides less error packets that leads to lower retransmissions and shorter delays.

Again referring to FIG. 7 & *8*, for the case of *a* = 1 our scheme gives a lower throughput compared to the uniform case where each mobile gets K/2 packets regardless of any parameters, but on the other side our scheme gives a lower PER because our scheme, adapts the forwarding process to the changes in PER of the links, lower PER means better performance, so our scheme surely outperforms the uniform case.

The variable K is an important parameter in our scheme, the performance of the whole system is related to it, large values of K will result in lower delays and higher throughput but with an increase in the PER. *Further investigation concerning the variable K will be included in the final version of this paper.*

## 5. Conclusion

In this paper, we proposed an approach for resource allocation at the relay node under data type constraints. This approach is based on a cross layer scheme which uses information, provided by the physical layer, at the MAC layer in order to achieve an efficient allocation that takes into account the Real-Time constraint for mobiles receiving Real-Time data and that provides Real-time user with an advantage over non-Real-Time user in terms of throughput and delay. We showed also that using a single physical parameter (like the PER) we can achieve the allocation goals without additional exchange of information.

**Bibliography**

**[1]** P. Liu, Z. Tao, Z. Lin, E. Erkip and S. Panwar "Cooperative wireless communications: a cross-layer approach" New York Polytechnic University.

**[2]** T. Korakis, Z. Tao, Y. Slutskiy, S. Panwar "A cooperative MAC protocol for Ad Hoc wireless networks" New York Polytechnic University.

**[3]** C. Chien, M. B. Srivastava, R. Jain, P. Lettieri, V. Aggarwal and R. Sternowski "Adaptive radio for Multimedia wireless Links " IEEE journal on selected areas in communications, Vol. 17, NO. 5, MAY 1999.

**[4]** H-Y. Wei, R. D. Giltin "Two-Hop-Relay architecture for Next-Generation WWAN/WLAN integration" Columbia University & NEC Laboratories America and Columbia University.

**[5]** IEEE 802.11 standard, Medium Access Control (MAC) and Physical layer (PHY) specifications. IEEE 2007